\documentclass[12pt, preprint]{emulateapj} % for a referee version
\usepackage{graphicx}
\usepackage{subfigure}
\begin{document}
\title{The June 2008 flare of Markarian 421 from optical to TeV energies\\}

\author{
I.~Donnarumma\altaffilmark{1,*}, V.~Vittorini\altaffilmark{1,2},
S.~Vercellone\altaffilmark{3}, E.~Del Monte\altaffilmark{1},
M.~Feroci\altaffilmark{1}, F.~D'Ammando\altaffilmark{1,2},
L.~Pacciani\altaffilmark{1}, A.~W. Chen\altaffilmark{3,4}, M.~Tavani\altaffilmark{1,2}, A.~Bulgarelli\altaffilmark{5},
A.~Giuliani\altaffilmark{3}, F.~Longo\altaffilmark{6},
G.~Pucella\altaffilmark{1}, A.~Argan\altaffilmark{1},
G.~Barbiellini\altaffilmark{6}, F.~Boffelli\altaffilmark{7},
P.~Caraveo\altaffilmark{3}, P.~W.~Cattaneo\altaffilmark{7},
V.~Cocco\altaffilmark{1}, E.~Costa\altaffilmark{1}, G.~De
Paris\altaffilmark{1}, G.~Di Cocco\altaffilmark{5},
Y.~Evangelista\altaffilmark{1}, M.~Fiorini\altaffilmark{3},
T.~Froysland\altaffilmark{2,4}, M.~Frutti\altaffilmark{1},
F.~Fuschino\altaffilmark{5}, M.~Galli\altaffilmark{8},
F.~Gianotti\altaffilmark{5}, C.~Labanti\altaffilmark{5},
I.~Lapshov\altaffilmark{1}, F.~Lazzarotto\altaffilmark{1},
P.~Lipari\altaffilmark{9}, M.~Marisaldi\altaffilmark{5},
M.~Mastropietro\altaffilmark{10}, S.~Mereghetti\altaffilmark{3},
E.~Morelli\altaffilmark{5}, A.~Morselli\altaffilmark{11},
A.~Pellizzoni\altaffilmark{3}, F.~Perotti\altaffilmark{3},
P.~Picozza\altaffilmark{11}, G.~Porrovecchio\altaffilmark{1},
M.~Prest\altaffilmark{12}, M.~Rapisarda\altaffilmark{13},
A.~Rappoldi\altaffilmark{7}, A.~Rubini\altaffilmark{1},
P.~Soffitta\altaffilmark{1}, M.~Trifoglio\altaffilmark{5},
A.~Trois\altaffilmark{1}, E.~Vallazza\altaffilmark{6},
A.~Zambra\altaffilmark{3}, D.~Zanello\altaffilmark{9},
C.~Pittori\altaffilmark{14}, P.~Santolamazza\altaffilmark{14},
F.~Verrecchia\altaffilmark{14}, P.~Giommi\altaffilmark{14},
S.~Colafrancesco\altaffilmark{14}, L.~Salotti\altaffilmark{15}}
\author{(The AGILE Team)}
\author{M.~Villata\altaffilmark{16,**},
C.~M.~Raiteri\altaffilmark{16,**}, W.~P.~Chen\altaffilmark{17},
N.~V.~Efimova\altaffilmark{18}, B.~Jordan\altaffilmark{19},
T.~S.~Konstantinova\altaffilmark{18},
E.~Koptelova\altaffilmark{17},
O.~M.~Kurtanidze\altaffilmark{20,21,22},
V.~M.~Larionov\altaffilmark{18,23},
%E.~Lindfors\altaffilmark{46}, %in MAGIC
J.~A.~Ros\altaffilmark{24}, A.~C.~Sadun\altaffilmark{25}}
%L.~O.~Takalo\altaffilmark{46}} %in MAGIC
%
\author{(The GASP-WEBT Team)}
\author{
H.~Anderhub\altaffilmark{26},
L.~A.~Antonelli\altaffilmark{27},
P.~Antoranz\altaffilmark{28},
M.~Backes\altaffilmark{29},
C.~Baixeras\altaffilmark{30},
S.~Balestra\altaffilmark{28},
J.~A.~Barrio\altaffilmark{28},
H.~Bartko\altaffilmark{31},
D.~Bastieri\altaffilmark{32},
J.~Becerra Gonz\'alez\altaffilmark{33},
J.~K.~Becker\altaffilmark{29},
W.~Bednarek\altaffilmark{34},
K.~Berger\altaffilmark{34},
E.~Bernardini\altaffilmark{35},
A.~Biland\altaffilmark{26},
R.~K.~Bock\altaffilmark{31,32},
G.~Bonnoli\altaffilmark{36},
P.~Bordas\altaffilmark{37},
D.~Borla Tridon\altaffilmark{31},
V.~Bosch-Ramon\altaffilmark{37},
T.~Bretz\altaffilmark{38},
I.~Britvitch\altaffilmark{26},
M.~Camara\altaffilmark{28},
E.~Carmona\altaffilmark{31},
A.~Chilingarian\altaffilmark{40},
S.~Commichau\altaffilmark{26},
J.~L.~Contreras\altaffilmark{28},
J.~Cortina\altaffilmark{39},
M.~T.~Costado\altaffilmark{33,41},
S.~Covino\altaffilmark{27},
V.~Curtef\altaffilmark{29},
F.~Dazzi\altaffilmark{32},
A.~De Angelis\altaffilmark{42},
E.~De Cea del Pozo\altaffilmark{43},
R.~de los Reyes\altaffilmark{28},
B.~De Lotto\altaffilmark{42},
M.~De Maria\altaffilmark{42},
F.~De Sabata\altaffilmark{42},
C.~Delgado Mendez\altaffilmark{33},
A.~Dominguez\altaffilmark{44},
D.~Dorner\altaffilmark{26},
M.~Doro\altaffilmark{32},
D.~Elsaesser\altaffilmark{38},
M.~Errando\altaffilmark{39},
D.~Ferenc\altaffilmark{45},
E.~Fern\'andez\altaffilmark{39},
R.~Firpo\altaffilmark{39},
M.~V.~Fonseca\altaffilmark{28},
L.~Font\altaffilmark{30},
N.~Galante\altaffilmark{31},
R.~J.~Garc\'{\i}a L\'opez\altaffilmark{33,41},
M.~Garczarczyk\altaffilmark{39},
M.~Gaug\altaffilmark{33},
F.~Goebel\altaffilmark{31,A},
D.~Hadasch\altaffilmark{30},
M.~Hayashida\altaffilmark{31},
A.~Herrero\altaffilmark{33,41},
D.~H\"ohne-M\"onch\altaffilmark{38},
J.~Hose\altaffilmark{31},
C.~C.~Hsu\altaffilmark{31,***},
S.~Huber\altaffilmark{38},
T.~Jogler\altaffilmark{31},
D.~Kranich\altaffilmark{26},
A.~La Barbera\altaffilmark{27},
A.~Laille\altaffilmark{45},
E.~Leonardo\altaffilmark{36},
E.~Lindfors\altaffilmark{46},
S.~Lombardi\altaffilmark{32},
M.~L\'opez\altaffilmark{32},
E.~Lorenz\altaffilmark{26,31},
P.~Majumdar\altaffilmark{35},
G.~Maneva\altaffilmark{47},
N.~Mankuzhiyil\altaffilmark{42},
K.~Mannheim\altaffilmark{38},
L.~Maraschi\altaffilmark{27},
M.~Mariotti\altaffilmark{32},
M.~Mart\'{\i}nez\altaffilmark{39},
D.~Mazin\altaffilmark{39},
M.~Meucci\altaffilmark{36},
M.~Meyer\altaffilmark{38},
J.~M.~Miranda\altaffilmark{28},
R.~Mirzoyan\altaffilmark{31},
J.~Mold\'on\altaffilmark{37},
M.~Moles\altaffilmark{44},
A.~Moralejo\altaffilmark{39},
D.~Nieto\altaffilmark{28},
K.~Nilsson\altaffilmark{46},
J.~Ninkovic\altaffilmark{31},
I.~Oya\altaffilmark{28},
R.~Paoletti\altaffilmark{36},
J.~M.~Paredes\altaffilmark{37},
M.~Pasanen\altaffilmark{46},
D.~Pascoli\altaffilmark{32},
F.~Pauss\altaffilmark{26},
R.~G.~Pegna\altaffilmark{36},
M.~A.~Perez-Torres\altaffilmark{44},
M.~Persic\altaffilmark{42,48},
L.~Peruzzo\altaffilmark{32},
F.~Prada\altaffilmark{44},
E.~Prandini\altaffilmark{32},
N.~Puchades\altaffilmark{39},
A.~Raymers\altaffilmark{40},
W.~Rhode\altaffilmark{29},
M.~Rib\'o\altaffilmark{37},
J.~Rico\altaffilmark{49,39},
M.~Rissi\altaffilmark{26},
A.~Robert\altaffilmark{30},
S.~R\"ugamer\altaffilmark{38},
A.~Saggion\altaffilmark{32},
T.~Y.~Saito\altaffilmark{31},
M.~Salvati\altaffilmark{27},
M.~Sanchez-Conde\altaffilmark{44},
P.~Sartori\altaffilmark{32},
K.~Satalecka\altaffilmark{35,***},
V.~Scalzotto\altaffilmark{32},
V.~Scapin\altaffilmark{42},
T.~Schweizer\altaffilmark{31},
M.~Shayduk\altaffilmark{31},
K.~Shinozaki\altaffilmark{31},
S.~N.~Shore\altaffilmark{50},
N.~Sidro\altaffilmark{39},
A.~Sierpowska-Bartosik\altaffilmark{43},
A.~Sillanp\"a\"a\altaffilmark{46},
J.~Sitarek\altaffilmark{31,34},
D.~Sobczynska\altaffilmark{34},
F.~Spanier\altaffilmark{38},
A.~Stamerra\altaffilmark{36},
L.~S.~Stark\altaffilmark{26},
L.~Takalo\altaffilmark{46},
F.~Tavecchio\altaffilmark{27},
P.~Temnikov\altaffilmark{47},
D.~Tescaro\altaffilmark{39},
M.~Teshima\altaffilmark{31},
M.~Tluczykont\altaffilmark{35},
D.~F.~Torres\altaffilmark{49,43},
N.~Turini\altaffilmark{36},
H.~Vankov\altaffilmark{47},
A.~Venturini\altaffilmark{32},
V.~Vitale\altaffilmark{42},
R.~M.~Wagner\altaffilmark{31,***},
W.~Wittek\altaffilmark{31},
V.~Zabalza\altaffilmark{37},
F.~Zandanel\altaffilmark{44},
R.~Zanin\altaffilmark{39},
J.~Zapatero\altaffilmark{30}}
\author{(The MAGIC Collaboration)}
\author{
V.~Acciari\altaffilmark{51,52},
E.~Aliu\altaffilmark{53},
T.~Arlen\altaffilmark{54},
M.~Beilicke\altaffilmark{55},
W.~Benbow\altaffilmark{51},
S.~M.~Bradbury\altaffilmark{56},
J.~H.~Buckley\altaffilmark{55},
V.~Bugaev\altaffilmark{55},
Y.~Butt\altaffilmark{57},
K.~Byrum\altaffilmark{58},
A.~Cannon\altaffilmark{59},
A.~Cesarini\altaffilmark{60},
Y.~C.~Chow\altaffilmark{54},
L.~Ciupik\altaffilmark{61},
P.~Cogan\altaffilmark{62},
P.~Colin\altaffilmark{63},
W.~Cui\altaffilmark{64},
M.~K.~Daniel\altaffilmark{56},
R.~Dickherber\altaffilmark{55},
C.~Duke\altaffilmark{65},
T.~Ergin\altaffilmark{57},
S.~J.~Fegan\altaffilmark{54},
J.~P.~Finley\altaffilmark{64},
G.~Finnegan\altaffilmark{63},
P.~Fortin\altaffilmark{66},
A.~Furniss\altaffilmark{67},
D.~Gall\altaffilmark{64},
G.~H. Gillanders\altaffilmark{60},
R.~Guenette\altaffilmark{62},
G.~Gyuk\altaffilmark{61},
J.~Grube\altaffilmark{56,59,****},
D.~Hanna\altaffilmark{62},
J.~Holder\altaffilmark{53},
D.~Horan\altaffilmark{58,76},
C.~M.~Hui\altaffilmark{63},
T.~Brian Humensky\altaffilmark{68},
A.~Imran\altaffilmark{69},
P.~Kaaret\altaffilmark{70},
N.~Karlsson\altaffilmark{61},
M.~Kertzman\altaffilmark{71},
D.~Kieda\altaffilmark{63},
J.~Kildea\altaffilmark{51},
A.~Konopelko\altaffilmark{72},
H.~Krawczynski\altaffilmark{55},
F.~Krennrich\altaffilmark{69},
M.~J. Lang\altaffilmark{60},
S.~LeBohec\altaffilmark{63},
G.~Maier\altaffilmark{62},
A.~McCann\altaffilmark{62},
M.~McCutcheon\altaffilmark{62},
A.~Milovanovic\altaffilmark{56},
P.~Moriarty\altaffilmark{52},
T.~Nagai\altaffilmark{69},
R.~A.~Ong\altaffilmark{54},
A.~N.~Otte\altaffilmark{67},
D.~Pandel\altaffilmark{70},
J.~S.~Perkins\altaffilmark{53},
A.~Pichel\altaffilmark{73},
M.~Pohl\altaffilmark{69},
K.~Ragan\altaffilmark{64},
L.~C.~Reyes\altaffilmark{74},
P.~T.~Reynolds\altaffilmark{75},
E.~Roache\altaffilmark{51},
H.~J.~Rose\altaffilmark{56},
M.~Schroedter\altaffilmark{69},
G.~H.~Sembroski\altaffilmark{64},
A.~W.~Smith\altaffilmark{58},
D.~Steele\altaffilmark{61},
S.~P.~Swordy\altaffilmark{68},
M.~Theiling\altaffilmark{51},
J.~A.~Toner\altaffilmark{60},
L.~Valcarcel\altaffilmark{62},
A.~Varlotta\altaffilmark{64},
S.~P.~Wakely\altaffilmark{68},
J.~E.~Ward\altaffilmark{59},
T.~C.~Weekes\altaffilmark{51},
A.~Weinstein\altaffilmark{54},
D.~A.~Williams\altaffilmark{67},
S.~Wissel\altaffilmark{68},
M.~Wood\altaffilmark{54},  
B.~Zitzer\altaffilmark{64}}
\author{(The VERITAS Collaboration)}

\altaffiltext{1} {INAF/IASF-Roma, I-00133 Roma, Italy} 
\altaffiltext{2} {Dip. di Fisica, Univ. Tor Vergata, I-00133 Roma, Italy}
\altaffiltext{3} {INAF/IASF-Milano, I-20133 Milano, Italy} 
\altaffiltext{4} {CIFS-Torino, I-10133 Torino, Italy} 
\altaffiltext{5} {INAF/IASF-Bologna, I-40129 Bologna, Italy}
\altaffiltext{6} {Dip. Fisica and INFN Trieste, I-34127 Trieste, Italy} 
\altaffiltext{7} {INFN-Pavia, I-27100 Pavia, Italy} 
\altaffiltext{8} {ENEA-Bologna, I-40129 Bologna, Italy} 
\altaffiltext{9} {INFN-Roma La Sapienza, I-00185 Roma, Italy} 
\altaffiltext{10} {CNR-IMIP, Roma, Italy} 
\altaffiltext{11} {INFN Roma Tor Vergata, I-00133 Roma, Italy}
\altaffiltext{12} {Dip. di Fisica, Univ. Dell'Insubria, I-22100 Como, Italy} 
\altaffiltext{13} {ENEA Frascati,  I-00044 Frascati (Roma), Italy}
\altaffiltext{14} {ASI Science Data Center, I-00044 Frascati(Roma), Italy} 
\altaffiltext{15} {Agenzia Spaziale Italiana, I-00198 Roma, Italy}
\altaffiltext{16}{INAF, Osservatorio Astronomico di Torino, Italy}
\altaffiltext{17}{Institute of Astronomy, National Central
University, Taiwan} \altaffiltext{18}{Astronomical Institute, St.
Petersburg State University, Russia} \altaffiltext{19}{School of
Cosmic Physics, Dublin Institute For Advanced Studies, Ireland}
\altaffiltext{20}{Abastumani Astrophysical Observatory, Georgia}
\altaffiltext{21}{Astrophysikalisches Institut Potsdam, Germany}
\altaffiltext{22}{Landessternwarte Heidelberg-K\"onigstuhl,
Germany} \altaffiltext{23}{Pulkovo Observatory, Russia}
\altaffiltext{24}{Agrupaci\'o Astron\`omica de Sabadell, Spain}
\altaffiltext{25}{Department of Physics, University of Colorado at
Denver, CO, USA}
\altaffiltext{26} {ETH Zurich, CH-8093 Switzerland}
\altaffiltext{27} {INAF National Institute for Astrophysics, I-00136 Rome, Italy}
\altaffiltext{28} {Universidad Complutense, E-28040 Madrid, Spain}
\altaffiltext{29} {Technische Universit\"at Dortmund, D-44221 Dortmund, Germany}
\altaffiltext{30} {Universitat Aut\`onoma de Barcelona, E-08193 Bellaterra, Spain}
\altaffiltext{31} {Max-Planck-Institut f\"ur Physik, D-80805 M\"unchen, Germany}
\altaffiltext{32} {Universit\`a di Padova and INFN, I-35131 Padova, Italy}
\altaffiltext{33} {Inst. de Astrof\'{\i}sica de Canarias, E-38200 La Laguna, Tenerife, Spain}
\altaffiltext{34} {University of \L\'od\'z, PL-90236 Lodz, Poland}
\altaffiltext{35} {Deutsches Elektronen-Synchrotron (DESY), D-15738 Zeuthen, Germany}
\altaffiltext{36} {Universit\`a  di Siena, and INFN Pisa, I-53100 Siena, Italy}
\altaffiltext{37} {Universitat de Barcelona (ICC/IEEC), E-08028 Barcelona, Spain}
\altaffiltext{38} {Universit\"at W\"urzburg, D-97074 W\"urzburg, Germany}
\altaffiltext{39} {IFAE, Edifici Cn., Campus UAB, E-08193 Bellaterra, Spain}
\altaffiltext{40} {Yerevan Physics Institute, AM-375036 Yerevan, Armenia}
\altaffiltext{41} {Depto. de Astrofisica, Universidad, E-38206 La Laguna, Tenerife, Spain}
\altaffiltext{42} {Universit\`a di Udine, and INFN Trieste, I-33100 Udine, Italy}
\altaffiltext{43} {Institut de Cienci\`es de l'Espai (IEEC-CSIC), E-08193 Bellaterra, Spain}
\altaffiltext{44} {Inst. de Astrof\'{\i}sica de Andalucia (CSIC), E-18080 Granada, Spain}
\altaffiltext{45} {University of California, Davis, CA-95616-8677, USA}
\altaffiltext{46} {Tuorla Observatory, Turku University, FI-21500 Piikki\"o, Finland}
\altaffiltext{47} {Institute for Nuclear Research and Nuclear Energy, BG-1784 Sofia, Bulgaria}
\altaffiltext{48} {INAF/Osservatorio Astronomico and INFN, I-34143 Trieste, Italy}
\altaffiltext{49} {ICREA, E-08010 Barcelona, Spain}
\altaffiltext{50} {Universit\`a  di Pisa, and INFN Pisa, I-56126 Pisa, Italy}
%%%%%%%
\altaffiltext{51}{Fred Lawrence Whipple Observatory, Harvard-Smithsonian Center for Astrophysics, Amado, AZ 85645, USA}
\altaffiltext{52}{Department of Life and Physical Sciences, Galway-Mayo Institute of Technology, Dublin Road, Galway, Ireland}
\altaffiltext{53}{Department of Physics and Astronomy and the Bartol Research Institute, University of Delaware, Newark, DE 19716, USA}
\altaffiltext{54}{Department of Physics and Astronomy, University of California, Los Angeles, CA 90095, USA}
\altaffiltext{55}{Department of Physics, Washington University, St. Louis, MO 63130, USA}
\altaffiltext{56}{School of Physics and Astronomy, University of Leeds, Leeds, LS2 9JT, UK}
\altaffiltext{57}{Harvard-Smithsonian Center for Astrophysics, 60 Garden Street, Cambridge, MA 02138, USA}
\altaffiltext{58}{Argonne National Laboratory, 9700 S. Cass Avenue, Argonne, IL 60439, USA}
\altaffiltext{59}{School of Physics, University College Dublin, Belfield, Dublin 4, Ireland}
\altaffiltext{60}{School of Physics, National University of Ireland, Galway, Ireland}
\altaffiltext{61}{Astronomy Department, Adler Planetarium and Astronomy Museum, Chicago, IL 60605, USA}
\altaffiltext{62}{Physics Department, McGill University, Montreal, QC H3A 2T8, Canada}
\altaffiltext{63}{Physics Department, University of Utah, Salt Lake City, UT 84112, USA}
\altaffiltext{64}{Department of Physics, Purdue University, West Lafayette, IN 47907, USA }
\altaffiltext{65}{Department of Physics, Grinnell College, Grinnell, IA 50112-1690, USA}
\altaffiltext{66}{Department of Physics and Astronomy, Barnard College, Columbia University, NY 10027, USA}
\altaffiltext{67}{Santa Cruz Institute for Particle Physics and Department of Physics, University of California, Santa Cruz, CA 95064, USA}
\altaffiltext{68}{Enrico Fermi Institute, University of Chicago, Chicago, IL 60637, USA}
\altaffiltext{69}{Department of Physics and Astronomy, Iowa State University, Ames, IA 50011, USA}
\altaffiltext{70}{Department of Physics and Astronomy, University of Iowa, Van Allen Hall, Iowa City, IA 52242, USA}
\altaffiltext{71}{Department of Physics and Astronomy, DePauw University, Greencastle, IN 46135-0037, USA}
\altaffiltext{72}{Department of Physics, Pittsburg State University, 1701 South Broadway, Pittsburg, KS 66762, USA}
\altaffiltext{73}{Casilla de Correo 67, Sucursal 28, (C1428ZAA) Ciudad Aut\`onoma de Buenos Aires, Argentina}
\altaffiltext{74}{Kavli Institute for Cosmological Physiscs, University of Chicago, Chicago, IL 60637, USA}
\altaffiltext{75}{Department of Applied Physics and Instumentation, Cork Institute of Technology, Bishopstown, Cork, Ireland}
\altaffiltext{76}{Laboratoire Leprince-Ringuet, Ecole Polytechnique, CNRS/IN2P3, F-91128, Palaiseau, France}
\altaffiltext{a}{deceased}

\altaffiltext{*}{AGILE Team Corresponding Author: I. Donnarumma,
immacolata.donnarumma@iasf-roma.inaf.it}
\altaffiltext{**}{GASP-WEBT Team Corresponding Author: M. Villata, villata@oato.inaf.it, C. M. Raiteri, raiteri@oato.inaf.it}
\altaffiltext{***}{MAGIC Collaboration Corresponding Author: R. M. Wagner, robert.wagner@mpp.mpg.de; C. C. Hsu, cchsu@mpp.mpg.de; K. Satalecka, satalk@ifh.de}
\altaffiltext{****} {VERITAS Team Corresponding Author: J. Grube, jeffrey.grube@ucd.ie}

\begin{abstract}
We present optical, X-ray, high energy ($\lessapprox 30$ GeV) and
very high energy ($\gtrapprox 100$ GeV; VHE) observations of the
high-frequency peaked blazar Mrk 421 taken between 2008 May 24 and June 23. 
A high energy $\gamma$-ray signal was detected by 
AGILE with $\sqrt{TS}=4.5$ on June 9--15, with 
$F(E>100~\mathrm{MeV})= 42^{+14}_{-12}\times 10^{-8}$ photons cm$^{-2}$ s$^{-1}$. 
This flaring state is brighter than the 
average flux observed by EGRET by a factor of 
$\sim$3, but still consistent with the highest EGRET 
flux. In hard X-rays (20-60 keV) SuperAGILE resolved a 5-day flare (June 9-15) 
peaking at $\sim$ 55 mCrab. SuperAGILE, RXTE/ASM and Swift/BAT data show 
a correlated flaring structure between soft and hard X-rays. Hints of 
the same flaring behavior are also detected in the simultaneous 
optical data provided by the GASP-WEBT. A Swift/XRT observation near the 
flaring maximum revealed the highest 2-10 keV flux ever observed from this 
source, of $2.6$ $\times 10^{-9}$ erg cm$^{-2}$ s$^{-1}$ (i.e. $> 100$ mCrab). 
A peak synchrotron energy of $\sim$3 keV was derived, higher than typical 
values of $\sim$0.5-1 keV. VHE observations with MAGIC and VERITAS on June 6-8
show the flux peaking in a bright state, well correlated with the X-rays. 
%We present the analysis of this extraordinary set of simultaneous
%data, covering a twelve-decade spectral coverage, in terms of the spectral energy distribution as well
%as of correlated light curves. The
This extraordinary set of simultaneous data, covering a twelve-decade 
spectral range, allowed for a deep analysis of the spectral energy 
distribution as well as of correlated light curves. The $\gamma$-ray flare 
can be interpreted within the framework of the synchrotron self-Compton 
model in terms of a rapid acceleration of leptons in the jet. 
\end{abstract}

\shorttitle{Observations of Mrk 421 in June 2008} 
\shortauthors{I. Donnarumma et al.}

\section{Introduction}

Mrk 421 is a nearby blazar ($z=0.031$) and one of the brightest BL Lac objects given its distance of 134.1 Mpc ($H_0=71$
km s$^{-1}$  Mpc$^{-1}$, $\Omega_{m}=0.27, \Omega_{\lambda}=0.73$) . It was observed in $\gamma$-rays by EGRET (Lin et
al. 1992) and it was the first extragalactic object detected at $E> 500$
GeV \cite{punch92}. It belongs to the class of High-energy peaked
BL Lac objects (HBLs) (Padovani \& Giommi 1995), i. e. radio-loud active
galactic nuclei with high radio and
optical polarization. Its spectral energy distribution (SED) is
double-humped with a first peak usually in the soft to medium
X-ray range, and a second one at GeV-TeV energies 
(Sambruna et al. 1996; Fossati et al. 1998). 
The first hump is commonly interpreted as due to synchrotron 
radiation from high-energy electrons in a relativistic jet, while 
the origin of the second peak is still uncertain. In the leptonic 
scenario it is interpreted as inverse Compton (IC) scattering of 
the synchrotron (Synchrotron self-Compton, SSC) or external 
photons (External Compton, EC)  by the same population of 
relativistic electrons. The observed correlated variability 
between X-rays and TeV $\gamma$-rays (Maraschi et al. 1999; Fossati et 
al. 2008, Wagner 2008) is well explained in the SSC framework 
(Ghisellini et al. 1998), whereas the EC scenario is unlikely to 
apply in HBLs, due to the low density of ambient photons. 
Alternatively, hadronic models invoke proton-initiated 
cascades and/or proton-synchrotron emission (Aharonian 2000, 
M\"{u}cke et al. 2003).

Leptonic and hadronic scenarios for HBLs predict different properties of the 
$\gamma$-ray emission in relation to emissions in other energy bands. 
Specifically, the hadronic models (as opposed to the SSC ones) predict 
a flatter slope of the $\sim 100$ MeV IC emission than that of the 
synchrotron emission in the optical-UV energy bands. $\gamma$-ray 
observations of flaring BL Lacs and simultaneous multiwavelength data 
are thus the keys to investigating these two scenarios.

A hard X-ray flare of Mrk 421 was detected by SuperAGILE on 2008 June 10
(Costa et al. 2008). This detection was later followed by a 
detection in $\gamma$-rays (Pittori et al. 2008) by the AGILE/GRID 
(Gamma Ray Imaging Detector) and prompted a ToO observation by Swift/XRT, 
complementing the ongoing multifrequency observing campaign of Mrk 421 with 
WEBT (optical), MAGIC and VERITAS (TeV). We report on the observations and 
the analysis of these data, complemented by the publicly-available data 
from RossiXTE/ASM (2-12 keV) and Swift/BAT (15-50 keV), and discuss the 
spectral energy distribution of the source during this bright $\gamma$-ray 
flare.

\section{Observations and Results}

\subsection{AGILE observations}
AGILE (Tavani et al. 2008) is a small mission of the Italian Space Agency for the exploration of the $\gamma$-ray sky, operating in a low
Earth orbit since 2007 April 23. The AGILE composite payload allows for
simultaneous observations in the energy ranges 30 MeV-30 GeV and 20-60 keV over
a very wide field of view by means of GRID and the hard X-ray imager
SuperAGILE, respectively. Mrk 421 was observed for five days, between 2008-06-09
17:02 UT and 2008-06-15 02:17 UT.

\subsubsection{Hard X-ray observations}
\label{SA}

%SuperAGILE is the hard X-ray monitor on-board AGILE operating in the nominal energy range 18-60 keV.
%It consists of four one-dimensional coded-mask detectors, encoding the same
%direction in pairs. The collimator and the mask define a single
%(one-dimensional) field of view of $107^{o} \times 68^{o}$ for each of the $4$
%detectors, orientated at $90^{o}$ for the two pairs of units.

On 2008 June 10 SuperAGILE detected enhanced hard X-ray emission from
Mrk 421. The measured flux in 20-60 keV was found to be above 30 mCrab 
(24-hour average), almost an order of magnitude larger than its typical 
flux in quiescence. The source was at $\sim (0,20)$ degrees in the orthogonal
SuperAGILE reference frame (Feroci et al. 2007), thus exposing an effective 
area $\sim55\%$ of the full on-axis value. 
%This assured us the detection of the source with one day
%integration time.
In the following days, the flux increased up to about 55 mCrab. 
The 5-day 20-60 keV SuperAGILE light curve is shown in Fig. 1c. 
%The flux increased by a
%factor of $\sim$2 over $\sim$4 days, starting to decrease the day
%before the end of the observation.
Using the publicly available light curves for this source from the BAT 
experiment\footnote[1]{http://swift.gsfc.nasa.gov/docs/swift/results/transients/weak/Mrk421/} in the 15-50 keV energy range (near the SuperAGILE bandpass), 
we calculated daily averages of the sparse observations.  This
allowed us to obtain good coverage also before and after the AGILE
observation (black squares in Fig. 1c), revealing that SuperAGILE indeed observed the maximum brightness of this hard X-ray flare. Both SuperAGILE 
and BAT count rates were converted to physical units by assuming 
a Crab-like spectrum.%, thus allowing a comparison of the two data
%set despite the slightly different bandpass.%

SuperAGILE photon-by-photon data allows extraction of a time-averaged 
energy spectrum from the mask-convolved
data. Given the lack of substantial spectral variability in the
SuperAGILE/ASM hardness ratio (Fig.1e) we accumulated the average energy
spectrum from the data of the last 4 days of the observations, when
the source flux varied between 35 and 55 mCrab, for a total net
source exposure of $\sim 140$ ks. Despite poor
statistics, this 4-point energy spectrum is able to reasonably constrain the photon index of a simple
power law, $\Gamma =2.43^{+0.69}_{-0.64}$
($\chi^{2}_{\rm dof}=0.8$, 2 dof). The average flux is $F(20-60 \mathrm {keV})
=(4.90 \pm 0.54) \times
10^{-10}$ erg cm$^{-2} \rm s^{-1}$ (($9.8 \pm 1.1) \times
10^{-3}$ photons cm$^{-2}$
s$^{-1}$).

\subsubsection{Gamma-ray Observations}

The GRID data were analyzed using the AGILE standard
pipeline (Vercellone et. al 2008), with a bin-size
of $0.25^{\circ} \times 0.25^{\circ}$. Only events flagged as $\gamma$-rays
and not recorded while the satellite crosses the South Atlantic Anomaly were accepted. We also rejected all 
events with reconstructed direction within $10^{\circ}$ from the
Earth limb thus reducing contamination from Earth's $\gamma$-ray albedo.

Mrk 421, observed $\sim$20$^{\circ}$ off-axis with respect to the boresight, was not detected on daily time scales, implying a
daily-averaged flux smaller than $\sim100 \times 10^{-8}$ photons
cm$^{-2}$ s$^{-1}$, similar to what observed by EGRET
\cite{har99}. A 4.5-$\sigma$ significance in
the range 100 MeV-10 GeV resulted from an integration over
the whole 5-day period (net exposure $\sim 260$ ks). The measured flux is  ($42^{+14}_{-12})
\times 10^{-8}$ photons cm$^{-2}$ s$^{-1}$, about $\sim$3 times
higher than the average flux detected by EGRET ($\sim13 \times
10^{-8}$, \cite{har99}) and $\sim$1.5 times higher than, but still
consistent, with the highest flux ($27\pm 7)\times 10^{-8}$ photons cm$^{-2}$
s$^{-1}$ observed during the Viewing Period 326. %(Both AGILE
%and EGRET uncertainties are given at 1-$\sigma$.) %This indicates
%an enhanced \gamma-ray activity, similarly to what observed at
%X-rays.
The short duration of the AGILE observation
did not allow us to follow the possible decay of the $\gamma$-ray
emission. Due to the relative faintness of the source and the short
exposure, the GRID data do not allow us to extract a reliable
energy spectrum.

\subsection{The Soft X-ray band}

\subsubsection{Swift X-ray telescope}

Following the SuperAGILE detection, on 2008 June 12 we triggered a
ToO observation with the {\it Swift} X-Ray Telescope (XRT) that
promptly observed the source for $~5$\,ks between 2008 June 12
19:33:20 UT (MJD 54629.8) and June 13 at 01:57:37 UT (MJD
54630.1).

The XRT data were processed using the standard
procedure ({\tt xrtpipeline} v0.12.0) developed at the ASI Science Data Center
and distributed by the NASA/HEASARC.
%, filtering and screening criteria by using FTOOLS in the {\tt Heasoft} package (v.6.4).
Observations were taken in Window Timing mode at a count rate of 120--150 cts s$^{-1}$. We selected XRT grades 0--2 (Burrows et al. 2005)
and extracted the WT data in a rectangular 40$\times$20-pixel
region centered on the source.
The background was also extracted within a box
(40$\times$20-pixel), far from background sources.

Due to calibration uncertainties, we restricted our spectral fit
to the energy range 0.7--9.0~keV and added a 3\% systematics to
the model (Cusumano et al. 2006). The XRT spectral data are well
described by an absorbed log-parabolic model. A joint fit of the XRT 
and SuperAGILE spectral data using the 4-day average spectrum described 
in Sect. 2.1.1. used the log-parabolic model with Galactic absorption ($N_{\rm H}^{\rm Gal} = 1.61 \times
10^{20}$\,cm$^{-2}$; Lockman \& Savage 1995), defined as:
\begin{center}
$F(E) = K E^{-a-b \log(E)}$  photons cm$^{-2}$ s$^{-1}$
\end{center}
where $a$ is the photon index at 1 keV and $b$ is the curvature
parameter (Massaro et al. 2004; Massaro et al. 2008). This
model usually describes adequately the featureless and curved
spectrum in HBLs.
The joint fit provides
$a=1.65_{-0.02}^{+0.01}$, $b=0.37_{-0.005}^{+0.01}$
($\chi^{2}_{\rm dof}$=1.4, 763 dof),  which implies  a peak energy
$2.97^{+0.22}_{-0.18}$ keV, and predicts $F_{\rm 2-10keV} =
2.56\times10^{-9}$\,erg\,cm$^{-2}$\,s$^{-1}$ (or $0.4$
photons\,cm$^{-2}$\,s$^{-1}$) and $F_{\rm 20-60keV}=
(5.7\pm 0.6)\times10^{-10}$\,erg\,cm$^{-2}$\,s$^{-1}$ (($1.1 \pm 0.1) \times 10^{-2}$
photons\,cm$^{-2}$\,s$^{-1}$), comparable to the
stand-alone SuperAGILE best fit.

\subsubsection{RossiXTE All Sky Monitor (ASM)}

Given the high flux observed by Swift/XRT, we retrieved the public
light curves provided by the
ASM\footnote[2]{http://xte.mit.edu/asmlc/ASM.html} to trace the evolution of
the soft X-rays during the AGILE observation.
Fig. 1b shows the daily light curve of Mrk 421
in the energy range 2-12 keV, obtained by properly averaging the
dwell-by-dwell data.

The emission at soft X-rays is well correlated with hard
X-ray emission. The ASM data show that the XRT observation took
place when the source was at its maximum emission at soft X-rays
(MJD $\sim$ 54630). Comparing the relative intensity
of the two flares in Fig. 1, it appears that the second peak is
significantly harder than the first one. This is also shown in Fig. 1e, 
where we computed the daily-averaged hardness ratio between hard (15--60 keV) 
and soft (2--12 keV) X-rays. The source appears to have undergone the hardest 
part of this double-humped flare just during the AGILE $\gamma$-ray detection.

\begin{figure}[hbt]
\begin{center}
%\subfigure{\rotatebox{90}{\includegraphics[width=0.45\textwidth]{magn_shr_surface_complete.ps}}}

%\includegraphics[width=7.5cm, height=15cm]{f1.eps}
\epsscale{0.8}
\plotone{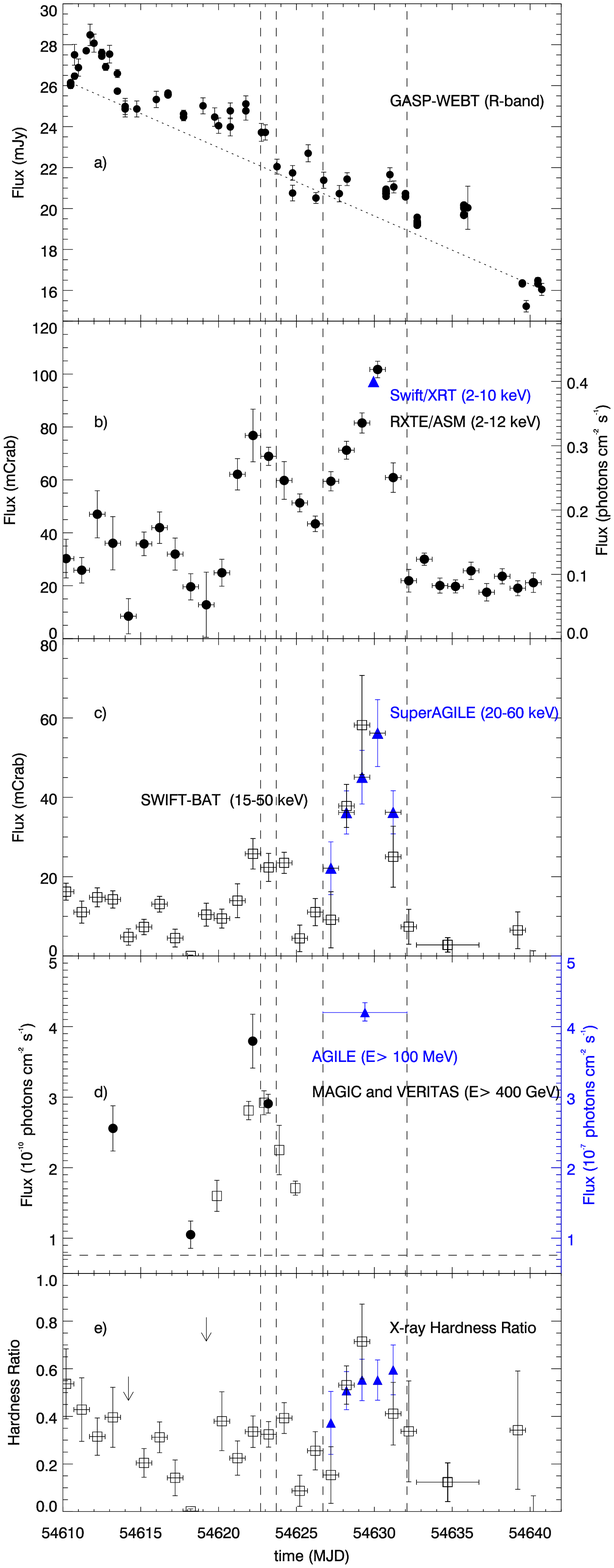}

\end{center}
 \caption{\footnotesize{\textbf{a)} $R$-band optical light curve from
GASP-WEBT (May 24--June 23); \textbf{b)} ASM (2-12 keV) light
   curve (bin size is 1 day) and XRT (2-10 keV) flux (blue triangle);
   \textbf{c)} SuperAGILE (20-60 keV, blue triangles; 1 Crab $= 0.2 $ ph cm$^{-2}$ s$^{-1}$)
   and BAT (15-50 keV, empty black squares; 1 Crab $= 0.29$ ph cm$^{-2}$
   s$^{-1}$); \textbf{d)} MAGIC and VERITAS ($>$400 GeV,
   empty black squares and black circles, respectively), the Crab
   flux $> 400$ GeV (horizontal dashed line), AGILE ($> 100$ MeV, blue triangle); \textbf{e)} the hardness ratio computed by using the
  SuperAGILE and ASM data for each day. The dashed vertical lines mark
  period 1 and period 2 (see text for details).}}

\end{figure}

\subsection{Observations in the  VHE $\gamma$-rays}

The 4-telescope array VERITAS (Acciari et al. 2008) at the Fred
Lawrence Whipple Observatory (Arizona) and the
single-dish instrument MAGIC (Baixeras et al. 2004; Tescaro et al. 2007) 
at La Palma are imaging air Cherenkov telescopes covering an energy range 
from $\sim 100$ GeV to some tens of TeV. The instruments have a typical 
energy resolution of $<$20\% (VERITAS) and 20--30\% (MAGIC), and 
event-by-event angular resolution of $<$ 0.14$^\circ$. Wobble-mode 
observations (Daum et al. 1997) at an 0.4$^\circ$ offset from the camera 
center were taken on five nights from 2008 June 3--8 with MAGIC at zenith 
angles (ZA) between 28$^\circ$ and 48$^\circ$ and on 4 nights (May 27, 
June 1, 5, 6) with VERITAS (wobble offset: 0.5$^\circ$) at ZA between 
32$^\circ$ and 40$^\circ$ during partial moon light conditions. The total 
live-time after applying quality selection is 2.95 and 1.17 hours with 
MAGIC and VERITAS, respectively. The data were analyzed using the MAGIC 
(Albert et al. 2008a) and VERITAS (Daniel et al. 2007; Acciari et al. 2008) 
standard calibration and analyses and image parameters (Hillas 1985). For 
VERITAS, the $\gamma$-ray direction and air shower impact parameter on the 
ground are reconstructed using the stereoscopic techniques in Hofmann et al.
(1999) and Krawczynski et al. (2006). Any $\gamma$-ray excess is derived
from the $\theta^2$ distribution, where $\theta$
represents the angular distance between the source position in the
sky and the reconstructed arrival position of the air shower. For
MAGIC, $\theta$ is estimated using the DISP method (Fomin et al.
1994). For VERITAS, 99.9\% of the background of cosmic-rays is rejected by 
using selection cuts on $\theta^2$, mean scaled width and
length, and by using quality cuts of each event.The MAGIC analysis
utilizes a random forest method (Albert et al. 2008b) to
discriminate the dominating background of hadronic cosmic-ray
events and for the energy estimation of the $\gamma$-ray events. The energy 
and effective area of each event is reconstructed from Monte Carlo 
simulations. The integral flux and energy spectrum of the source is then 
derived from the effective areas for nights with a significant detection. 
Two independent analyses of both the MAGIC and VERITAS data sets yielded 
consistent results. In total, a signal corresponding to a significance level 
of $44\sigma$ (VERITAS) and $66\sigma$ (MAGIC) is obtained by following 
eq. 17 of Li \& Ma (1983). The combined MAGIC-VERITAS data (Fig. 1d) show a 
transient peaking near MJD 54622. The VERITAS 
energy spectrum for June 6 is provided. A power law fit over the energy range 
0.3--5 TeV resulted in a $\chi^{2}_{\rm dof}$=0.7 with a photon index 
$\Gamma = 2.78 \pm 0.09$. In Fig. 2 we show the intrinsic $\gamma$-ray spectrum at the source
reconstructed by removing attenuation effects by the extragalactic 
background light (Hauser \& Dwek 2001) following the procedure of Raue \& 
Mazin (2008) (\footnote[3]{http://www.desy.de/$\sim$mraue/ebl/}). Fitting a power law to the intrinsic spectrum yields a photon 
index $\Gamma = 2.59 \pm 0.08$, which is not significantly harder than the 
measured spectrum due to the relatively low redshift z $=$ 0.031.

\subsection{Optical and UV observations}

Mrk 421 is one of the 28 $\gamma$-ray-loud blazars that are
regularly monitored by the GLAST-AGILE Support Program
(GASP; Villata et al. 2008) of the Whole Earth Blazar Telescope
(WEBT)\footnote[4]{{\tt http://www.oato.inaf.it/blazars/webt}\\ see
e.g.\ \citep*{vil07, rai07}}. The GASP observations
started in early 2007 November. The $R$-band data were calibrated
according to the photometric sequence by Villata et al. (1998). A careful
data analysis is warranted because the source flux is contaminated
by the emission of the host galaxy and nearby objects. The flux
contribution of these objects was subtracted according to Nilsson et al. (2007). Moreover, we corrected for a Galactic extinction of
$A_R=0.042$ mag. The resulting $R$-band light curve\footnote[5]{The
optical data presented in
  this paper are stored in the GASP-WEBT archive; for questions regarding
  their availability, please contact the WEBT President Massimo Villata {\tt
    villata@oato.inaf.it}.} during the period considered in this paper is shown in
Fig.~1a.

During the Swift pointing on 2008 June 12--13, the UVOT (Poole et al. 2008) instrument observed Mrk 421 in the UVW1 and UVW2 photometric bands. The UVOTSOURCE tool is used to extract counts, correct for coincidence losses, apply background subtraction, and calculate the source flux. We applied a standard 5 arcsec radius source aperture, and a 20 arcsec background region. The source fluxes are dereddened using the interstellar extinction curve in Fitzpatrick (1999).

\section{Discussion}

Mrk 421 showed a very interesting broad-band activity
during the first half of 2008 June as derived from AGILE data
combined with those of WEBT, Swift, MAGIC and VERITAS.
Using our multi-frequency data we were able to derive
time-resolved SEDs (Fig. 2). We distinguish two time periods: {\it period 1}: 2008 June
6, with the inclusion of optical, X-ray (XTE and BAT) and TeV data (VERITAS); and {\it period 2}:
2008 June 9--15, including optical, UV, X-rays (XRT and SuperAGILE) and gamma-ray data (AGILE).
The source shows a very interesting time-variable broad band emission that appears to be in
overall agreement with an SSC model. 
%A more comprehensive
%account of the theoretical modelling will be presented elsewhere. appears to
%be in 
The optical, soft and hard X-ray bands strongly constrain the SED around 
the synchrotron peak, and its daily variability reveals the
physical processes of Mrk~421. Possible correlated variability is shown in Fig.~1 between the optical (an overall decreasing
trend with superimposed spikes of emission), the X-rays (several
emission peaks lasting few days), and the high-energy parts of the
spectrum. Based on the physical constraints obtained for the
synchrotron peak, we can model both the HE and VHE $\gamma$-ray 
emission. The data collection and broad-band SED extends over 12 decades in 
energy. Taking advantage of the overlapping MAGIC and VERITAS observations, 
we present a combined VHE light curve using the current 
generation of Northern imaging air Cherenkov systems.

%The VHE emission probed through the TeV light curve allowed us to 
%study in detail both synchrotron and IC emissions.  

% We note that due to moon
%constraints, TeV data are available only during the time period
%5-8 June, 2008.

We first model the synchrotron peak of emission using the period 1
optical, soft and hard X-ray data. The short
time-variability (Fig.~1) constrains the size of the emitting region to
$R<cT\delta\sim 5\times 10^{16}(\delta/20)$ cm. Hence,  we consider a one-zone SSC
model (Tavecchio et al. 1998) based on a blob of comoving size
$R=4\times 10^{16}$ cm, with a relativistic Doppler factor $\delta=20$  and
characterized by non-thermal comoving electron energy
distribution function described by a double power law:
\begin{equation}
n_{e}(\gamma)=\frac{K\gamma_{b}^{-1}}{(\gamma
/\gamma_{b})^{p_1}+(\gamma /\gamma_{b})^{p_2}}
\end{equation}
where the comoving Lorentz factor ($\gamma$) varies in the range  $\gamma_{min}=4 \times 10^{3}<\gamma<\gamma_{max}=1.3\times 10^{6}$, the normalization (density)
constant  $K=4\times 10^{-4}$ cm$^{-3}$, and the break energy $\gamma_{b}=
3.6\times 10^{5} $ and
with $p_1=2.22$, $p_2=4.5$, the low-energy and high-energy power-law
indices, respectively (see Table 1).  With these parameters we found that the data
for period 1  are best fitted with a comoving magnetic field $B=0.1$
G.

Variability may be caused by several factors; we consider two
cases: (A) hardening/softening of the electron energy
distribution function caused by particle acceleration processes;
(B) increase/decrease of the comoving particle density,
as a consequence of additional particle injection/loss by shock
processes.

We expect TeV variability to be comparable with the X-ray one if case
(A) applies: this is because the emission is in the Klein-Nishina
regime. Alternately, for case (B) we expect the TeV relative variability
($\Delta F/F$) to be a
factor of 2 greater than that of the X-ray flux variability.
%
%  spectra if the emission at TeV energies  due to
%  \mathbf{pure} enhancement of electron energies (as suggested by
%  the enhancement of the hardness ratio). But if the variability was
%  due to a density increase then the TeV flux would match the large
%  variability at soft X-rays.

Our AGILE, MAGIC and VERITAS data appear to support case (A). We
compare the SEDs for period 1 and period 2, to
better assess the spectral evolution.
In Fig.~2 we show our optimized modelling of the time-resolved synchrotron peak and consequent SSC high-energy emission for the period 1 as well as for 2008 June 12--13 of the period 2.
In the last case, the adopted model parameters are $p_1=2.1$, $p_2=5$,
$\gamma_{b}=4.2\times 10^{5}$, $K=6\times 10^{-4}$ cm$^{-3}$. 
Our model predicts an even
larger TeV flux for period 2 (no TeV observations exist, however) than
detected in period 1. 
%
% We note that the time variable
%TeV emission detected during both periods 1 and 2 is in the
%Klein-Nishina regime.
%
%
% Indeed, in Fig. 2 we report the SED also for June 6, 7, 8. Data in
% optical and X-ray bands are directly converted from the relevant
% light-curves assuming the same spectral indices as in June 12-13,
% TeV data are from VERITAS \mathbf{for June 6 and from MAGIC for
% June7-8}).
% As expected, these SEDs could be modelled as in June 12-13 but
% with a softer electron energy density: $\gamma_{b}=2.6\,10^{5}$,
% $p1=2.2$.

A detailed discussion of the complex optical vs. X-ray variability 
of Mrk~421 as shown in Fig.~1 will be presented elsewhere. We 
notice here a few remarkable points. The optical light curve shows 
variations of the order of $10\%$ on a time scale $\sim$few 
days, superimposed on a long decay during the entire period. 
Individual soft and hard X-ray peaks result in increased fluxes by a 
factor of $\sim$2.5 and $\sim$5, respectively: no long term decay appears. This different behaviour of 
the X-ray radiation and the bulk of the optical emission may interestingly 
suggest a more complex scenarios than A) and B) ones: optical and X 
radiation comes from two different jet regions, each one characterized 
by its own variability. A possible scenarios is one in which the inner 
jet region would produce the X-rays and it would be at least partially 
transparent to the optical radiation. 
In contrast, the outer region can only produce lower-frequency 
emission. The signature of the X-ray events visible in the optical 
light curve would come from the inner region and would be diluted 
by the optical radiation emitted from the outer region (see Villata \& Raiteri 
1999 for the case of Mrk 501; Villata et al. 2004).

Interestingly, the 2-10 keV flux measured by XRT on June 12 --
13,  $\sim 2.6\times 10^{-9}$ erg cm$^{-2}$ s$^{-1}$, is higher
than all previous observations ($< 2 \times 10^{-9}$ erg cm$^{-2}$
s$^{-1}$; Fossati et al. 2008; Lichti et al. 2008). A joint
analysis of the XRT and SuperAGILE data, covering the range from
0.7 to 60 keV, provides a best-fit spectral model consistent with
a log-parabolic shape, with parameters implying a peak energy $\sim$ 3 keV, in good agreement with the steeper
positive correlation between the peak energy and the maximum of
the SED found by Tramacere et al. 2007 (see their Fig.~3),
although our value of the peak energy shows a significant shift
with respect to typical values of 0.5-1 keV for this source.
% A
% time-resolved analysis of the XRT data shows further interesting
% features. The X-ray flux in the 0.3-2 keV and 2-10 keV energy
% ranges shows a gradual and continuous increase over the $\sim$6
% hours of the observation, passing from $\sim 2.2$ to $\sim 2.9
% \times 10^{-9}$ erg cm$^{-2}$ s$^{-1}$ in 2-10 keV and from $\sim
% 2.0$ to $\sim 2.3 \times 10^{-9}$ erg cm$^{-2}$ s$^{-1}$ in 0.3-2
% k eV. Remarkably, near the central time of the observation, the
% flux in both bands drops drastically, to $\sim 1.9$  and to $\sim
% 1.8 \times 10^{-9}$ erg cm$^{-2}$ s$^{-1}$, respectively. Such a
% short timescale variability is reminiscent of Mrk 501 at TeV
% energies (Albert et al. 2007, ApJ 669, 862). In contrast, the
% energy spectrum, as described by the parameter $a$, does not show
% significant variation over the first $\sim$4 hours, while in the
% last hour of the observation evidence for hardening is detected.

We conclude our analysis of the broad-band variable emission from
Mrk~421 by emphasizing that our multi-telescope/instrument data
show a very interesting variability that provides support for an
SSC model of the source. The $\gamma$-ray emission detected by
AGILE during period 2 and the TeV emission detected during
period 1 can be successfully modelled from the characteristics of
the corresponding synchrotron peaks.
%Our detection of optical,
%X-ray and gamma-ray flux during period-2 leads to a prediction of
%a TeV emission during that period even larger than that detected
%during period-1.

\begin{figure}[hbt]
\begin{center}
\epsscale{1.3}
\plotone{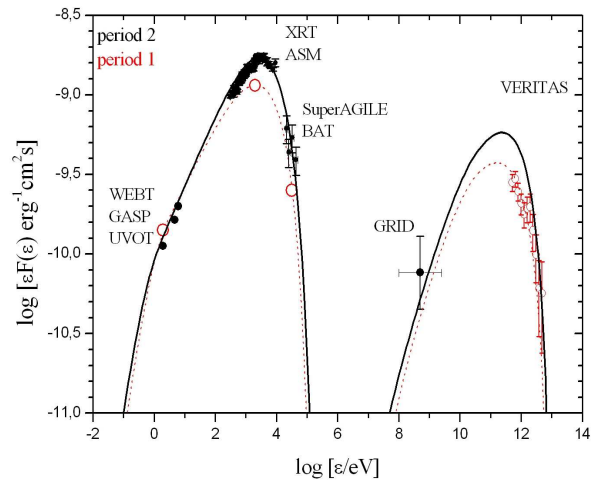}

\caption{SEDs of Mrk 421
obtained by combining the GASP-WEBT, SWIFT/UVOT, RossiXTE/ASM, XRT,
SuperAGILE, BAT, GRID and VERITAS data in period 1 and period 2 (red empty cirles and
black filled circles, respectively). Both  are
one-zone SSC models (red dashed line for period 1 and black solid line for
period 2).}
\end{center}
\end{figure}

%\begin{figure}[h!]
%\begin{center}
%\includegraphics[width=0.9\textwidth]{XRT-flux.ps}
%\caption{Swift-XRT light curve for the observation of June 12-13 (red circles
%  for 0.3-2 keV, black circles for 2-10 keV \textbf{Due to the Letters page-limit, this will be removed in the
%final version}}
%\end{center}
%\end{figure}

%\begin{figure}[h!]
%\begin{center}
%\includegraphics[width=0.9\textwidth]{XRT-index.ps}
%\caption{Spectral evolution across the XRT observation, as
%represented by the $a$ parameter of the log-parabolic model.
%\textbf{Due to the Letters page-limit, this will be removed in the
%final version}}
%\end{center}
%\end{figure}

%\begin{figure}[hbt]
%\begin{center}
%\rotatebox{-90}{\includegraphics[width=0.7\textwidth]{mrk421_comspe.ps}}
%\caption{Joint fit of the Swift-XRT and SuperAGILE spectra with log-parabolic
%  model: \textbf{Due to the Letters page-limit, this will be removed in the final version}}
%\end{center}
%\end{figure}

\acknowledgements
\footnotesize{AGILE is a mission of ASI, with
co-participation of INAF and
INFN. This work was
partially supported by ASI grants I/R/045/04, I/089/06/0, I/011/07/0
and by the Italian Ministry of University and Research (PRIN
2005025417), (ASDC) I/024/05/1. The MAGIC collaboration thanks the Instituto
de Astrofisica de Canarias for the excellent working conditions at the
Observatorio del Roque de Los Muchachos in La Palma and support by the German BMBF and MPG, the
Italian INFN and Spanish MCINN is acknowledged. This work was supported by ETH Research Grant TH 34/043, by the Polish MNiSzW
Grant N N203 390834, the YIP of the Helmholtz Gemeinschaft, the grant of
Georgian National Science Foundation GNSF/ST07/4-180. E. K. acknowledges
financial support from the NCS grant No. 96-2811-M-008-058.
The VERITAS collaboration is supported by grants from the U.S. 
Department of Energy, the National Science Foundation, and the 
Smithsonian Institution, by NSERC in Canada, Science Foundation 
Ireland, and PPARC in the UK. We acknowledge the technical support 
staff at the FLWO. We also acknowledge the Swift Team for carrying 
out the ToO observation.
}

\begin{table}
\caption{SSC model parameters.}
\centering
\begin{tabular}{ccc}
\\
\hline\hline
  parameter      &    period 1     & period 2 \\
\hline   
 $\gamma_b$      & $3.6 \cdot 10^5  $  & $4.2 \cdot 10^5$ \\
 $\gamma_{max}$  & $1.3 \cdot 10^6$  & $1.3 \cdot 10^6$ \\
 $p_1$           & 2.22               & 2.1 \\
 $p_2$           & 4.5              & 5 \\
 $B$ [G]         & 0.1               & 0.1 \\
 $K$ [cm$^{-3}$] & $4 \cdot 10^{-4} $ & $6 \cdot 10^{-4} $ \\
 $\delta$        & 20                & 20 \\
 $\theta[^{\circ}]$  & 2                 &2 \\         
\hline
\end{tabular}
\end{table}

{}
\end{document}